\documentclass{webofc}
\usepackage[varg]{txfonts}   
\begin{document}
\title{Higher Twists}
%
%

\author{\firstname{Vladimir M.} \lastname{Braun}
         \inst{1}
         \fnsep\thanks{\email{vladimir.braun@ur.de}} }

\institute{Institut f\"ur Theoretische Physik, Universit\"at
   Regensburg,  D-93040 Regensburg, Germany 
          }

\abstract{%
   The higher twist corrections refer to 
   a certain class of contributions to hard processes in strong interactions 
   that are suppressed by a power of the hard scale. 
  This is a very broad field of research which is becoming more and more 
  important as the accuracy of the available experimental data increases. 
  I give an overview of some relevant basic theory concepts and technical developments, 
  and briefly discuss a few phenomenological applications.  
}
\maketitle
\section{Introduction}
\label{intro}
The notion of twist was introduced in 1971 in the paper by Gross and Treiman \cite{Gross:1971wn} 
who noticed that ``it is no longer the dimension alone that determines the importance of an operator near the light-cone,
but rather the difference between the dimension and spin''. They called this quantity the ``twist'' on an operator, 
$\tau = d - s$. For a paradigm example of the traceless and symmetric in all indices quark-antiquark operator 
with many derivatives
\begin{align}
 O_{\mu\mu_1\ldots \mu_n} = \bar q \gamma_{\mu} D_{\mu_1}\ldots D_{\mu_n} q\,, && \tau = (n+3)-(n+1) = 2\,.
\end{align}
Contributions of the operators with the lowest twist give the dominant contribution to light-cone dominated processes and
those of higher twist (HT) are power-suppressed.
A similar observation was done around the same time by Brandt and Preparata \cite{Brandt:1970kg}.

Today, the name ``twist'' is used more broadly as a basis for several distinct classification schemes.
The original definition $\tau = \text{dimension}-\text{spin}$ is sometimes referred to as ``geometric twist''.
Operators with different geometric twist do not mix under renormalization and their matrix elements define 
independent nonperturbative quantities. In contrast, ``collinear twist'' is defined as 
$\tau = \text{dimension}-\text{spin~projection~on~plus~direction}$. This concept goes back to the light-cone formalism
by Kogut and Soper \cite{Kogut:1969xa} who decompose the quark and gluon fields 
in ``good'' (dynamically independent) and ``bad'' components. Each replacement of a ``good'' by a ``bad'' component adds
one unit of collinear twist. The utility of this concept is that collinear twist counting is directly related,
see e.g. \cite{Jaffe:1991ra}, to power suppression of the corresponding contribution in light-cone dominated processes, 
i.e. those that can be treated using collinear factorization. The price to pay is that parton distributions of a given 
collinear twist contain contributions of lower geometric twists which are dynamically independent. The classical example of such 
``contamination'' is provided by the Wandzura-Wilczek contribution \cite{Wandzura:1977qf} to the structure function $g_2(x,Q^2)$ 
which will be discussed in what follows. 

The subject of HT effects is getting prominence fuelled by very high accuracy of the  experimental data
from LHC, JLAB, KEKII, and on the 10 years scale from the Electron-Ion Collider (EIC) \cite{AbdulKhalek:2021gbh}. 
Interpretation of these data requires the corresponding theory precision so that higher-twist corrections cannot be ignored. 
Maybe more importantly, as emphasized already by Politzer \cite{Politzer:1980me}, the  higher-twist contributions
provide one with insight on  quark-gluon correlations and quantum interference effects in hadrons.
Thus their study allows one to learn more about hadron structure.

HT contributions do not have any simple partonic interpretation. Generally speaking, they are 
generated by a non-vanishing parton off-shellness/transverse momentum in two-particle parton distributions
and the contributions of multiparton correlation functions that take into account coherent hard scattering 
from a parton pair. These two effects are physically distinct, but are related by exact QCD equations of motion (EOM)
so that they must be taken into account simultaneously.
A good example is provided by the twist-three light-cone distribution amplitudes (LCDAs) of the $\pi$-meson,
schematically 
\begin{align}
  &\langle 0 | \bar q (z) \gamma_5 q(0)|\pi(q)\rangle  &&\mapsto \phi_p(x)
\notag\\
  &\langle 0 | \bar q (z) \sigma_{+-}\gamma_5 q(0)|\pi(q)\rangle  &&\mapsto \phi_\sigma(x)
\notag\\
  &\langle 0 | \bar q (z) \sigma_{+\perp}g F_{+\perp}(\alpha z) \gamma_5 q(0)|\pi(q)\rangle  &&\mapsto \Phi_{3}(x_1,x_2,x_3)
\end{align}
that contribute at subleading $1/Q$ accuracy in hard processes involving pion emission with large momentum.
One can show \cite{Braun:1989iv} that the two-particle LCDAs $\phi_p(x)$, $\phi_\sigma(x)$  are given by exact relations in terms of the three-particle LCDA $\Phi_{3}(x_1,x_2,x_3)$. 

Thanks to EOM there are several possibilities to identify independent degrees of freedom by choosing a convenient operator basis.
The most natural choice is probably a multipartonic (longitudinal)  basis in which case contributions of ``bad'' field components are 
systematically eliminated in terms of the contributions with extra gluon fields (or quark-antiquark pairs). This is the closest 
one can do to maintain partonic interpretation and Lorentz covariance. This approach was followed 
in most of the early works on higher twists, 
see \cite{Politzer:1980me,Shuryak:1981kj,Shuryak:1981pi,Jaffe:1982pm,Jaffe:1983hp,Bukhvostov:1984rns,Bukhvostov:1985rn}.
A ``transverse'' basis \cite{Ellis:1982cd}, where HT operators are built of a quark-antiquark pair and transverse covariant 
derivatives presents another viable option that simplifies diagrammatic calculation of the coefficient functions.  
An example of the application of this technique and its equivalence to the covariant approach 
can be found in \cite{Anikin:2009bf}. Worth mentioning is also the $SL(2)$-covariant operator basis of 
Refs.~\cite{Braun:2008ia,Braun:2009vc} that makes explicit symmetry properties and greatly simplifies the 
evolution equations.  The one-loop evolution equations for all twist-three and twist-four operators in QCD are 
known~\cite{Shuryak:1981kj,Shuryak:1981pi,Bukhvostov:1984rns,Bukhvostov:1985rn,Braun:2009vc}. Some of these equations 
are completely integrable \cite{Braun:1998id} and can be solved explicitly, see \cite{Belitsky:2004cz} for a review.  

\section{Twist three}\label{sec:t3}

\begin{figure*}[t]
\begin{center}
\includegraphics[width=0.40\textwidth]{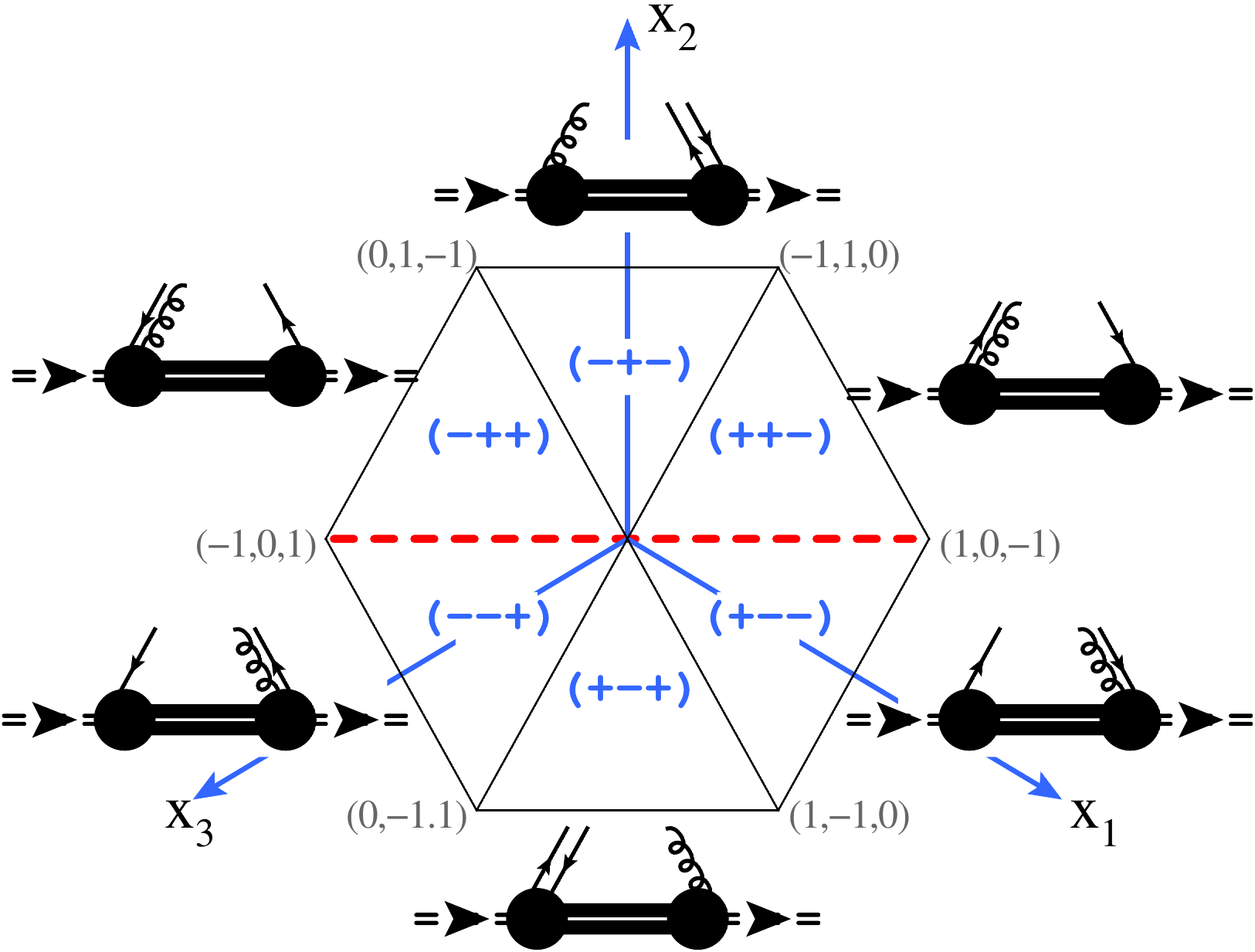}
\hspace{2em}
\includegraphics[width=0.40\textwidth]{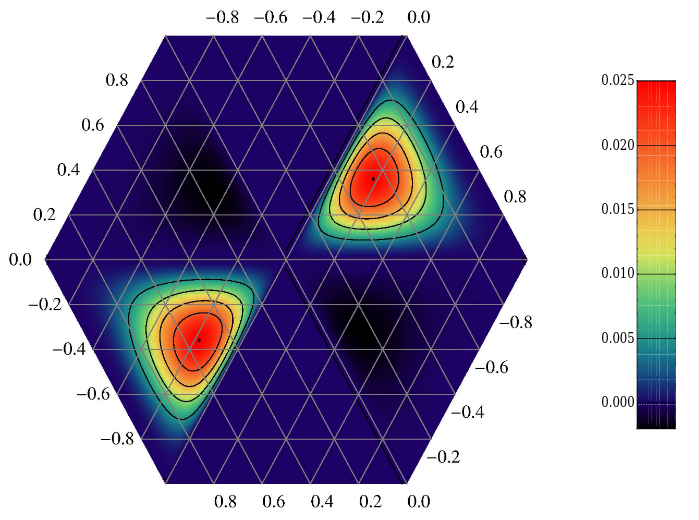}
\end{center}
\caption{Domain interpretation (left) and the light-front wave function overlap model \cite{Braun:2011aw} (right) for the 
twist-three $u$-quark-antiquark-gluon correlation function in the proton.}
\label{fig:hexagon}       
\end{figure*}

Twist-three contributions in inclusive and semiinclusive reactions in collinear factorization can be described in terms 
of a quark-antiquark-gluon correlation function (CF) (and in addition a three-gluon CF, for flavor singlet). 
It is a function of two variables and is conveniently presented in barycentric coordinates $x_1+x_2+x_3=0$ where 
$x_1$, $x_2$ and $-x_3$ correspond to the momentum fractions carried by the quark, gluon and antiquark, respectively,
see Fig.~\ref{fig:hexagon}.
This CF describes quantum-mechanical interference between scattering from a parton pair and a single parton, with six different
regions as shown  on the left panel.  
A simple light-front wave function overlap model \cite{Braun:2011aw}  for the u-quark  CF is shown on the right panel as an illustration.  
The scale-dependence of the twist-three CFs is well 
understood \cite{Bukhvostov:1984rns,Bukhvostov:1985rn,Balitsky:1987bk,Ji:1990br,Koike:1994st,Braun:2009mi}. A summary of the one-loop evolution 
kernels can be found in \cite{Braun:2009mi}.

The information on twist-three CFs can be obtained in particular from measurements of the structure function $g_2(x,Q^2)$ in polarized deep-inelastic scattering (DIS) and 
single transverse spin asymmetries in semininclusive reactions. In the first case a certain integral over the whole hexagon in 
Fig.~\ref{fig:hexagon} enters, and in the second case the CF on the horizontal line $x_2=0$ gives the main contribution, 
the so-called Qiu-Sterman function \cite{Qiu:1991pp}. Proposals and exploratory studies exist of various twist-three matrix elements in lattice 
QCD~\cite{Gockeler:2005vw,Bhattacharya:2020cen,Bhattacharya:2021moj,Burger:2021knd,Braun:2021aon}.

The structure function $g_2(x,Q^2)$ provides a paradigm case of a {\it collinear twist}-three observable (suppressed as $1/Q$) that receives contributions of
{\it geometric twist}-two and -three.  
\begin{align}
  & g_2^{(\tau 2)}(x,Q^2) =  g_1 (x,Q^2) - \int_x^1\!\frac{dy}{y} g_1 (y,Q^2) && \Leftarrow~~\text{Wandzura-Wilczek~contribution}
\notag\\
  & g_2^{(\tau 3)}(x,Q^2) \stackrel{!}{=}  D (x,Q^2) - \int_x^1\!\frac{dy}{y} D (y,Q^2) && \Leftarrow~~\text{convenient~parametrization}
\label{eq:g2}
\end{align} 
The first equation is exact in QCD, the second one is just a convenient parametrization. The D-function is given by a certain integral of the quark-antiquark-gluon
CF sketched in Fig.~\ref{fig:hexagon}, see e.g. \cite{Braun:2011aw}.    

\begin{figure*}[ht]
\begin{center}
\includegraphics[width=0.33\textwidth]{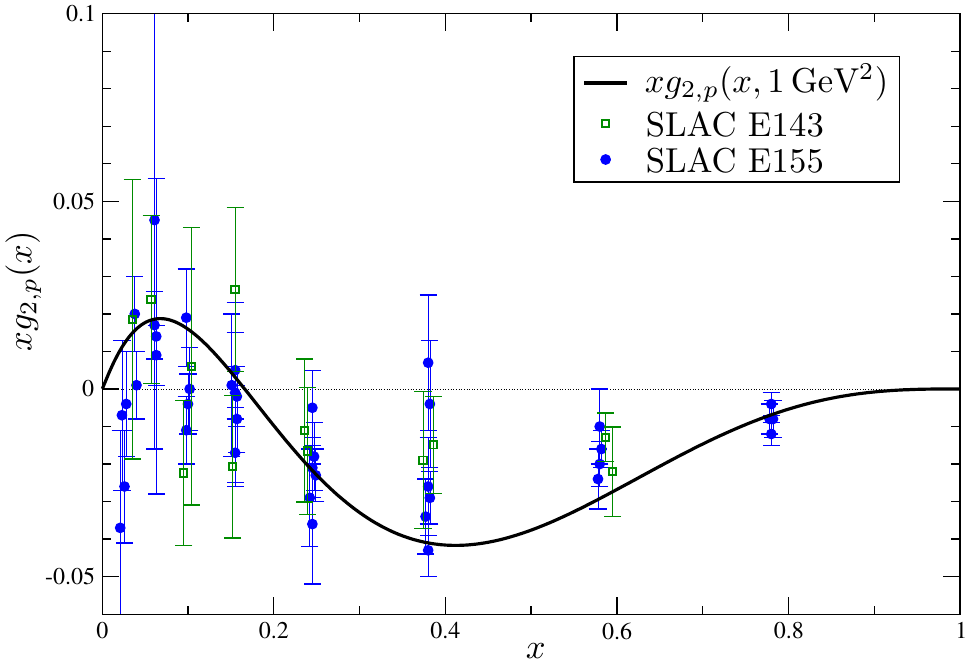}
\includegraphics[width=0.33\textwidth]{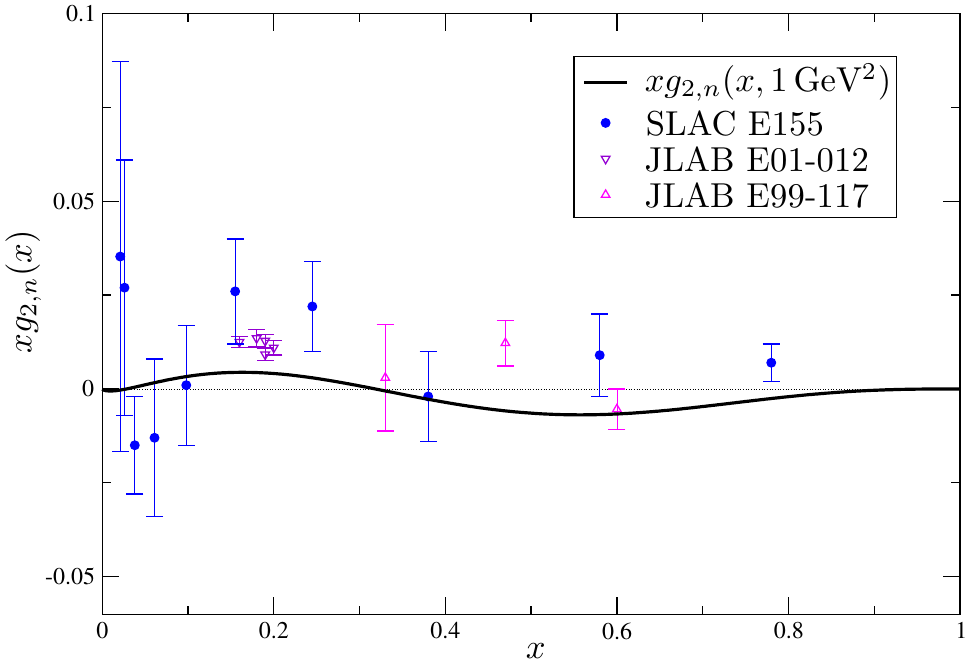}
\includegraphics[width=0.32\textwidth]{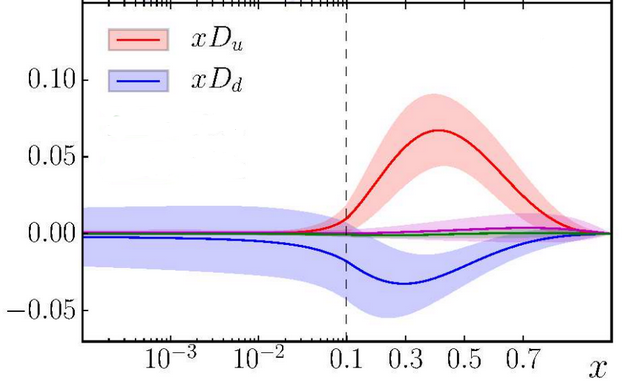}
\end{center}
\caption{ Left and middle panels: The structure function $g_2(x,Q^2)$ for the proton and the neutron, respectively.
           Right panel: The twist-three contributions $D(x,Q^2)$ for the $u$-quark and the $d$-quark in the proton. Figure adapted from 
\protect\cite{Sato:2016tuz}.}
\label{fig:JAM}       
\end{figure*}

Experimental data on the structure function $g_2(x,Q^2)$ so far are not very precise, see the first two panels in Fig.~\ref{fig:JAM}.  
The Wandzura-Wilczek contribution is known to be dominant, which makes the extraction of the twist-three part difficult.
A recent attempt \cite{Sato:2016tuz} is shown in Fig.~\ref{fig:JAM} on the right panel. 

\noindent
Better data exist for the first nontrivial moment $d_2 = 6 \int dx x^2\,g_2^{(\tau 3)}(x) $. 
A recent lattice calculation~\cite{Burger:2021knd} obtains  
$ d^{(p)}_2 = 0.0105(68)$ and $ d^{(n)}_2 = -0.0009(70)$ for the proton and the neutron, respectively. 
In this work also a rather complete compilation of the experimental data on $d_2$ can be found.
The smallness of $d_2$ was expected and supported by older estimates based on QCD sum rules \cite{Balitsky:1989jb,Stein:1994zk} 
and the instanton liquid model \cite{Balla:1997hf}.

\begin{figure}[t]
\begin{center}
\begin{minipage}{0.32\textwidth} 
\begin{center}
\includegraphics[width=0.99\textwidth,clip = true]{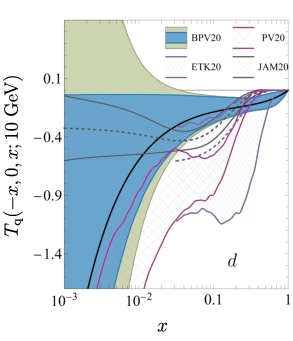}
\end{center}
\end{minipage}
\hspace*{5mm}
\begin{minipage}{0.33\textwidth} 
\begin{center}
\vspace*{-0.34cm}
\includegraphics[width=0.85\textwidth,clip = true]{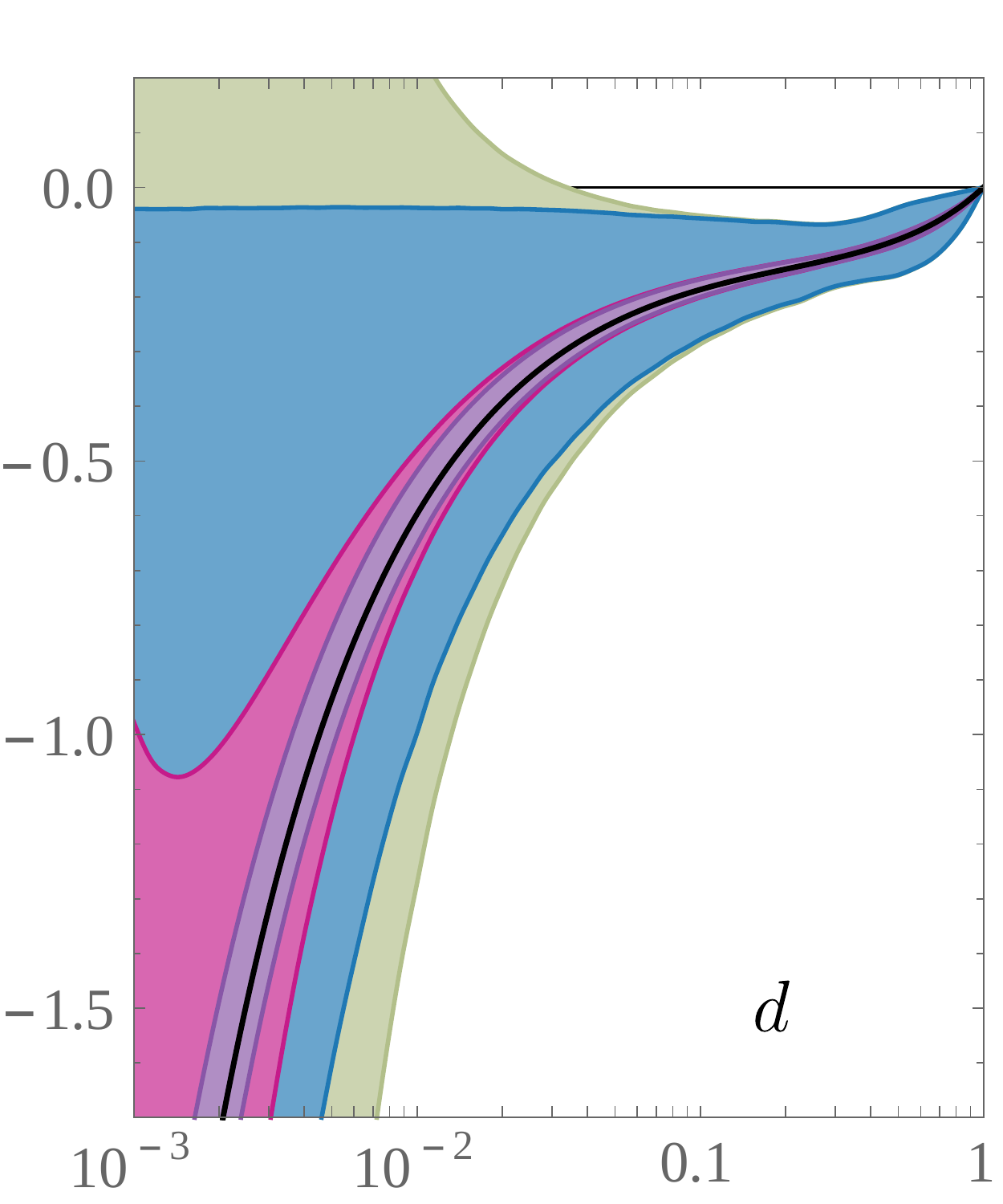}
\end{center}
\end{minipage}
\end{center}
\caption{The $d$-quark contribution to the Qiu-Sterman function extracted from the collinear limit of the Sivers function.
The existing determinations are summarized on the left panel (adapted from  \cite{Bury:2020vhj}). The projected accuracy for EIC
is shown in magenta on the right panel (EIC impact studies, courtesy of A. Vladimirov).}
\label{fig:QSfunction}
\end{figure}

Another source of information on the twist-three quark-antiquark-gluon correlation function is provided by the 
Qiu-Sterman function~\cite{Qiu:1991pp}. It can in principle be extracted, e.g., from the collinear limit of the Sivers function, but 
the results are so far very uncertain, see the left panel in Fig.~\ref{fig:QSfunction}. The situation is going to improve significantly
with the data from EIC, see the right panel for an impact study. The projected accuracy is indicated by the magenta region.

Due to the time constraints I will not discuss here subleading twist-three LCDAs of pseudoscalar and vector mesons. This is a large separate topic
where the formalism is well-developed and realistic models based on the conformal expansion \cite{Braun:2003rp} are available, see e.g. 
\cite{Braun:1989iv,Ball:1998sk,Ball:2006wn}.

\section{Twist four and renormalons}
\label{sec-3}

The structure of twist-four contributions to DIS is 
well-known since many years ~\cite{Shuryak:1981kj,Shuryak:1981pi,Jaffe:1982pm,Jaffe:1983hp,Bukhvostov:1984rns,Bukhvostov:1985rn,Ellis:1982cd,Balitsky:1987bk}.
A generalization to one-particle inclusive production in $e^+e^-$-annihilation also exists~\cite{Balitsky:1990ck}.   
The one-loop evolution equations for all twist-four correlation functions are known~\cite{Shuryak:1981kj,Shuryak:1981pi,Bukhvostov:1985rn,Braun:2009vc}.
The main problem is that the nonperturbative input required for the description of twist-four effects is very complicated; for the DIS case 
one finds seven independent three-parton and four-parton distributions (for one quark flavor) \cite{Jaffe:1983hp} and there is no obvious hierarchy.
The phenomenology is inconclusive; it is generally accepted that twist-four corrections in DIS are ``small'', but is statement is difficult to quantify.%
\footnote{In the case of twist-four LCDAs a systematic expansion is possible in conformal spin, which is analogous to the partial wave expansion in quantum 
mechanics. This construction is widely used to build models, see e.g.~\cite{Braun:1989iv,Ball:1998ff,Ball:2007zt}, 
with many applications to weak exclusive B-decays.}
A crucial difference to twist-three effects is that the twist-four contributions are not associated with particular power-suppressed {\it observables},
but rather give rise to $1/Q^2$ corrections to the amplitudes/cross sections that receive the leading-twist contributions as well. 

A separation of the leading power/leading twist and $1/Q^2$ twist-four contributions is in fact a nontrivial problem. The reason is that twist-four 
operators suffer from quadratic UV divergences so that their determination from, e.g., lattice QCD would need a prescription how these divergences are 
regularized. This prescription, in turn, affects the procedure to treat higher-order perturbative contributions at leading power due to the so-called 
renormalons. I will briefly explain this phenomenon and argue that the ``renormalon problem'' naturally leads to ``renormalon models'' for the power
corrections with very few parameters. A detailed discussion can be found in \cite{Beneke:1998ui,Beneke:2000kc}.      

The key observation is that a leading-twist calculation ``knows''  about the necessity to add a power correction.
Consider the DIS structure function $F_2(x,Q^2)$ as an example. It can be written as a convolution 
\begin{align}
 F_2(x,Q^2) &= 2x \int\limits_x^1 \frac{dy}{y} C(y,Q^2/\mu^2) q\Big(\frac{x}{y},\mu^2\Big)\left(1 + \frac{ D_2(x)}{Q^2}\right)\,.
\end{align}
Here $ C(y) = \delta(1-y) + \sum_{n=0}^\infty c_n \alpha_s^{n+1}$, $\alpha_s=\alpha_s(\mu)$,
$q(x,\mu^2)$ is the quark distribution function (PDF) and $D_2(x)$ is a twist-four contribution
of interest.

Imagine the separation between the coefficient function (CF) and the quark PDF is done using explicit cutoff at $|k|=\mu$.
The CF will then be modified compared to usual calculation by terms $\sim \mu^2/Q^2$:
\begin{align} 
   C_2(y)|^{\rm cut} &= \delta(1-y) + \sum_{n=0}^\infty c_n \alpha_s^{n+1} - \frac{\mu^2}{Q^2} d_2(x) 
  + \mathcal{O} \left( \frac{\mu^4}{Q^4}\right)\,. 
\label{cutoff}
\end{align} 
The dependence on $\mu$ must cancel: Logarithmic terms $\ln Q^2/\mu^2$ in CFs against the $\mu$-dependence in PDFs, and
power-suppressed terms $\mu^2/Q^2$ against the higher-twist contributions.
This implies  that $D_2(x)$ {\it in the cutoff scheme} must have the form
\begin{align}
   D_2(x) &= \mu^2 2x \int_x^1 \frac{dy}{y} d_2(x) q\left(\frac{x}{y}\right) +\delta D_2(x)\,,
\end{align}
where the first term is related to quadratic UV divergences in matrix elements of twist-4 operators (in this scheme).

Using dimensional regularization power-like terms in the CFs do not appear. 
Instead, the coefficients $c_k$ diverge factorially with the order $k$.  The sum of 
the pert. series in the CF is only defined to a power accuracy and this ambiguity (renormalon ambiguity) 
must be compensated by adding a non-perturbative higher-twist correction.
A detailed analysis~\cite{Beneke:2000kc} shows that  
large-order behavior of the coefficients (the renormalons) is in one-to-one correspondence 
with the sensitivity to extreme (small or large) loop momenta in Feynman diagrams. 
IR renormalons in twist-two  CFs are compensated by UV renormalons in matrix elements of twist-four operators. At
the end the same picture re-appears: only the details depend on the factorization method.

\begin{figure}[t]
\begin{center}
 \begin{minipage}[c]{5cm}\includegraphics[width=5cm, clip = true]{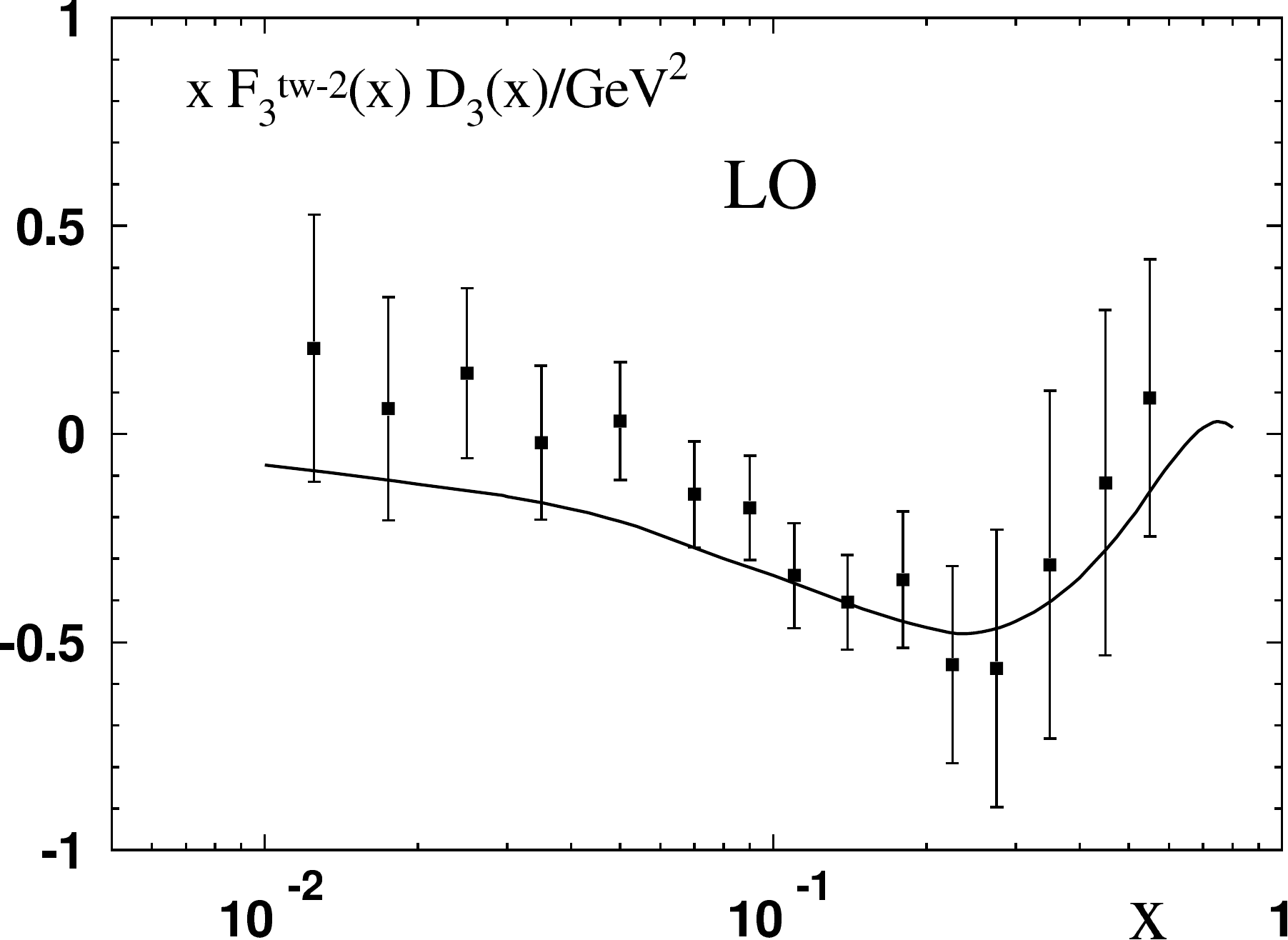}\end{minipage}
\hspace*{0.5cm}
 \begin{minipage}[c]{5cm}\includegraphics[width=5cm, clip = true]{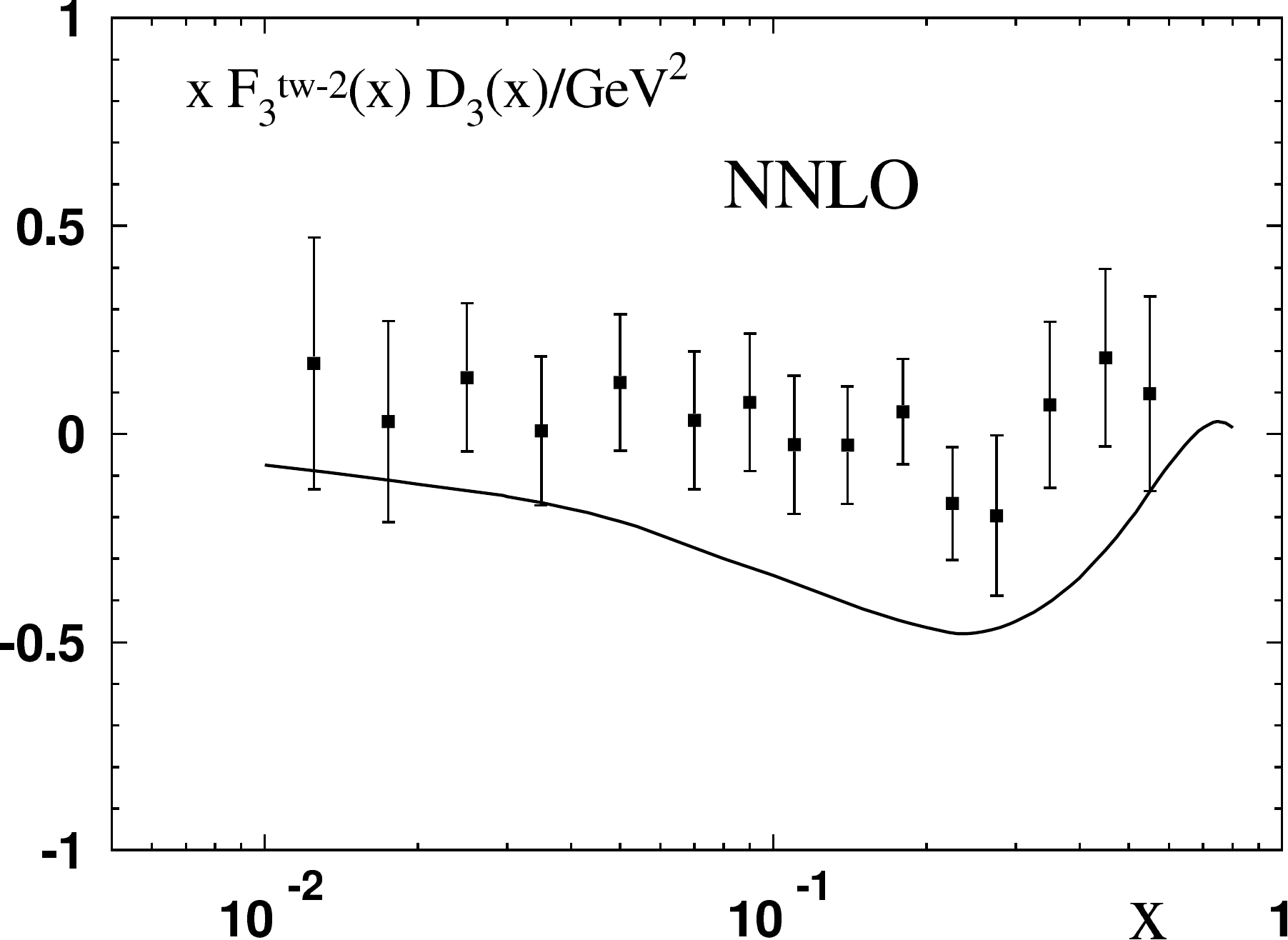}\end{minipage}
\end{center}
\caption{Twist-4 correction to $x F_3(x,Q^2)$ as extracted from the (revised) CCFR data. 
The data points \cite{Kataev:1997nc} are overlaid with the shape obtained from the ``renormalon model'' for the $1/Q^2$ power correction.
Figure adapted from \cite{Beneke:2000kc}.}
\label{fig:renormalons}
\end{figure}

The quadratic term in $\mu$ in Eq.~(\ref{cutoff}) is spurious since its sole purpose is to cancel the
similar contribution to the CF $\Rightarrow$ does not contribute to any physical observable.
Assuming, however, that the ``true'' twist-four term is of the same order, one gets a {\it renormalon model}  
\begin{align}
   D_2(x) &= \varkappa\Lambda^2_{\rm QCD} 2x \int\limits_x^1 \frac{dy}{y} d_2(x) q(\tfrac{x}{y}) \,, 
\qquad \varkappa = \mathcal{O}(1)\,,
\end{align}
where $\varkappa$ is the only free parameter.
To one-loop accuracy \cite{Dasgupta:1996hh}
\begin{align}
  d_2 &= -\frac{4}{[1-x]_+} + 4 + 2x + 12 x^2 - 9\delta(1-x) -\delta'(1-x)\,.
\end{align}
Noticeable features of this approximation is that the shape of the twist-four correction
does not depend on the target, and that the correction is enhanced at $x\to 1$:  
$D_2(x)/Q^2 \sim \Lambda^2/[Q^2(1-x)]$.
This analysis was done in the past for $F_2(x,Q^2)$, $F_L(x,Q^2)$, $F_3(x,Q^2)$ and $g_1(x,Q^2)$ etc.  
An example is shown in Fig.~\ref{fig:renormalons} where the data points \cite{Kataev:1997nc}
on the twist-four correction to $x F_3(x,Q^2)$ are overlaid with the shape obtained from the renormalon model.
This analysis also confirms the expectation from renormalons that the size of the twist-four correction 
is decreasing with the increasing order of perturbation theory in leading twist.

The power of the renormalon-based analysis is that this technique can be used for Minkowski-space observables where the operator product 
expansion is not applicable ---- the Drell-Yan process, event shapes in $e^+e^-$ annihilation etc., see  \cite{Beneke:1998ui,Beneke:2000kc}
for a review. Notable other applications are to power corrections in quasi- and pseudo-parton distributions \cite{Braun:2004bu,Braun:2018brg} 
and in transverse momentum dependent (TMD) distributions \cite{Scimemi:2016ffw}.
For a recent review on renormalons in the heavy quark pole mass definition see \cite{Beneke:2021lkq}.  

\section{Kinematic power corrections in off-forward reactions}

For definiteness, I will speak of the deeply-virtual Compton scattering (DVCS)
$\gamma^\ast(q) + N(p) \to \gamma(q') + N(p')$, $q^2 = -Q^2 \gg \Lambda^2_{\text QCD}$, $q'^2=0$, but the problem is generic for all
hard processes involving a momentum transfer between the initial and final state hadrons.
This reaction involves three helicity amplitudes --- helicity conserving and helicity flip for transversely polarized photons, 
and with a longitudinally polarized photon in the initial state --- of which the first one contributes at leading twist and
the other two are power suppressed. As well known, the definition of helicity amplitudes depends on the frame of reference.
Going over to a different frame, the leading-twist amplitude gets modified by power corrections
$(\sqrt{-t}/Q)^n$ and $(m/Q)^n$ where $t$ is the Mandelstam momentum transfer variable and $m$ is the nucleon mass, 
and also the subleading power amplitudes are generated even if they were absent (neglected) in the original frame.

\begin{figure}[t]
\begin{center}
\begin{minipage}[c]{3.5cm}\includegraphics[width=3.5cm, clip = true]{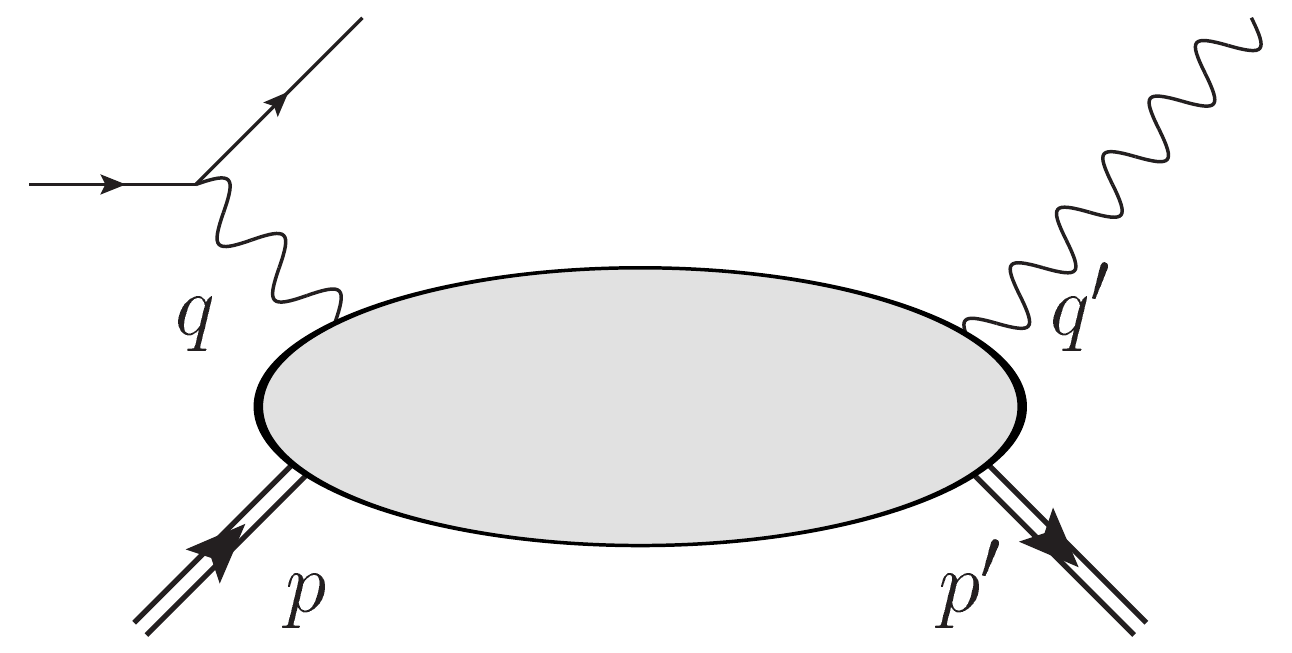} \end{minipage}
\begin{minipage}[l]{4cm}
{\small
\begin{align*}
 & \hspace*{-0.2cm}{\text{\footnotesize "DIS frame"}}
\\
   p &= (p_0, {\vec{0}_\perp}, p_z)
\\
   q &= (q_0, {\vec{0}_\perp}, q_z)
\end{align*}
}
\end{minipage}
\begin{minipage}[l]{4cm}
{\small
\begin{align*}
 & \hspace*{-0.2cm}{\text{\footnotesize "Photon frame"}}
\\
   q'&= (q'_0, {\vec{0}_\perp}, q'_z)
\\
   q &= (q_0,  {\vec{0}_\perp}, q_z)
\end{align*}
}
\end{minipage}
\end{center}
\caption{Two examples of possible choices of longitudinal and transverse directions in DVCS}
\label{fig:kin}
\end{figure}

\begin{figure}[t]
\begin{center}
{}\hspace*{3mm}\includegraphics[width=0.850\textwidth, clip]{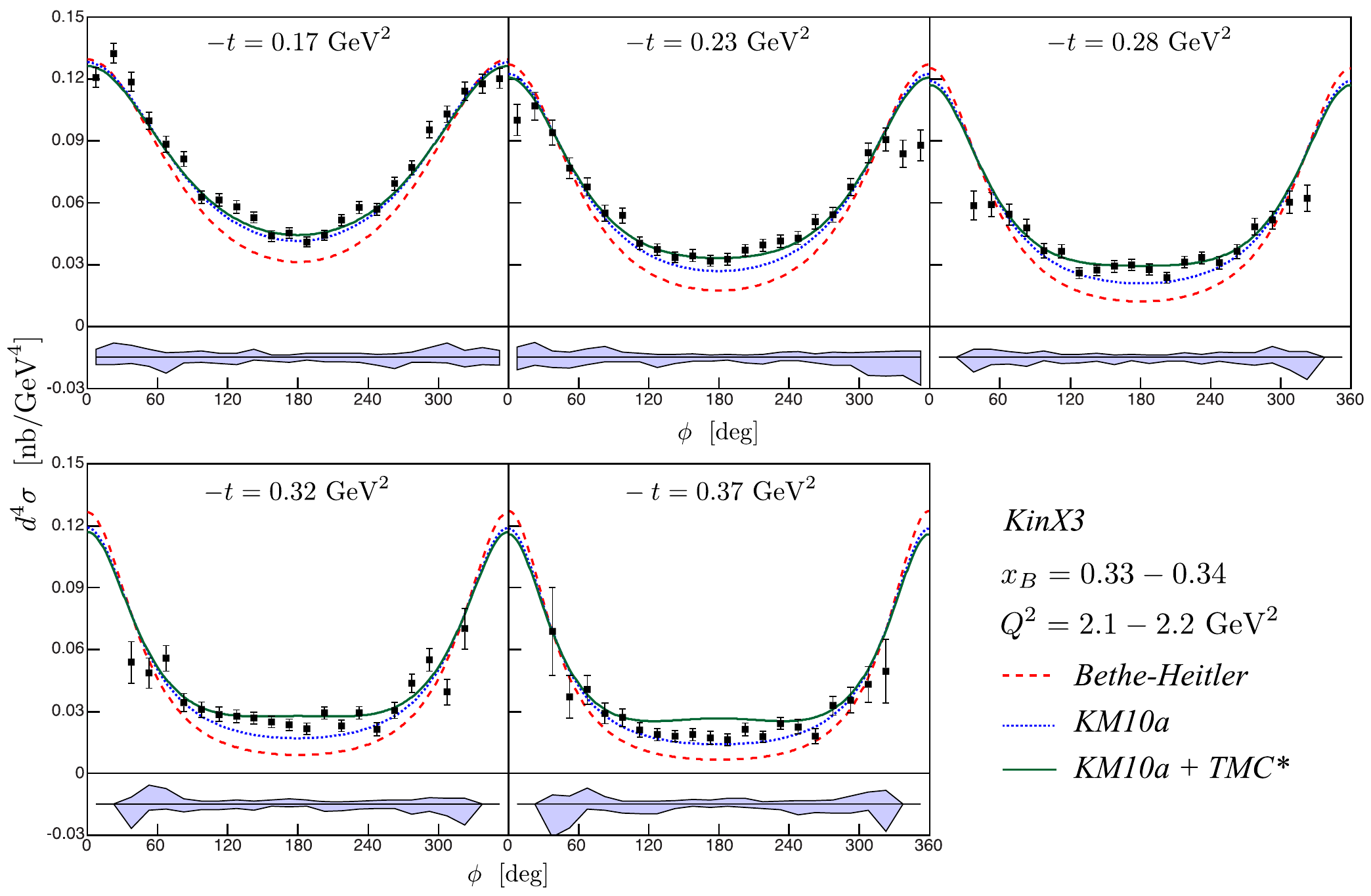}
\end{center}
\caption{The (unpolarized) DVCS cross section as a function of the azimuthal angle $\phi$ \cite{JeffersonLabHallA:2015dwe}
compared to the GK12 GPD model \cite{Kumericki:2009uq}.}  
\label{fig:HallA}
\end{figure}

Which frame is the best? 
In DIS there is a natural choice: 
the photon-nucleon scattering plane (in four dimensions) is defined as longitudinal and orthogonal directions 
as transverse. In DVCS the four external momenta are not complanar and there are many possibilities, 
see Fig.~\ref{fig:kin}.    
The leading-twist approximation in DVCS is therefore convention-dependent: taking into account only the leading-power contribution 
in the ``DIS frame'' and the ``photon frame'' will produce different results for the observables 
\cite{Braun:2014sta,Braun:2014paa,Guo:2021gru}. The difference can be large in certain kinematic regions, see Fig.~\ref{fig:HallA}.
The Bethe-Heitler contribution (red dashes) is essentially an electromagnetic background. The blue curve
shows the leading-twist result in the ``DIS frame'' using the GK12 GPD model \cite{Kumericki:2009uq}.
The green curve includes in addition the kinematic power corrections $t/Q^2$ and $m^2/Q^2$. 
This curve is very close numerically to the leading-twist result in the ``photon frame'', see~\cite{Braun:2014sta}. 
One sees that the difference is quite substantial.

Power corrections that repair the frame dependence (and restore electromagnetic gauge invariance) in DVCS  come from
two sources. First, there are corrections $(\sqrt{-t}/Q)^n$ and $(m/Q)^n$ to the matrix elements
of twist-two operators that are analogous to the target mass corrections in DIS \cite{Nachtmann:1973mr}. Second, there are
power corrections due to contributions of higher-twist operators that are obtained from the twist-two ones by adding total derivatives,
which are called the descendants. Schematically
\begin{align}
 \text{T}\{j(x) j(0)\} &= 
\sum_N \Big\{ A_N^{\mu_1\ldots \mu_N} \underbrace{\mathcal{O}^{N}_{\mu_1\ldots \mu_N}}_{\text{twist-2~operators}} +
  B_N^{\mu_1\ldots \mu_N} \underbrace{\partial^{\mu}\mathcal{O}^{N}_{\mu,\mu_1\ldots\mu_N}}_{\text{descendants~of~twist~2}} 
\notag\\&\qquad
+ 
 C_N^{\mu_1\ldots \mu_N} \underbrace{\partial^2\mathcal{O}^{N}_{\mu_1\ldots\mu_N}}_{\text{descendants}}
+
 D_N^{\mu_1\ldots \mu_N} \underbrace{\partial^{\mu}\partial^\nu\mathcal{O}^{N}_{\mu,\nu,\mu_1\ldots\mu_N}}_{\text{descendants}} +\ldots  \Big\}
\notag\\&\qquad 
+~\text{quark-gluon operators}
\end{align}

\noindent
Contributions of the descendants have to be taken into account, whereas the additional quark-antiquark-gluon (and more complicated)
operators can be omitted for the present purpose as they involve new nonperturbative matrix elements and cannot be responsible
for the restoration of Lorentz invariance.  
The problem is that matrix elements of some of the descendant operators on free quarks vanish, so that the corresponding coefficient
functions cannot be calculated in the standard manner.
Moreover, descendant operators are related to the $\bar qFq$ operators by EOM, for the simplest case
\begin{align}
  \partial^\mu O_{\mu\nu} = 2i\bar q g F_{\nu\mu}\gamma^\mu q\,, &&  O_{\mu\nu} = (1/2)[\bar q \gamma_\mu\!
  \stackrel{\leftrightarrow}{D}_\nu \!q
  + (\mu\leftrightarrow\nu)]\,.
\end{align}
Thus the separation of ``kinematic'' higher-twist contributions of the descendant operators from the ``genuine'' higher-twist
corrections due to quark-gluon correlations poses a nontrivial problem \cite{Braun:2011zr,Braun:2011dg}.
The suggested solution (see also \cite{Braun:2020zjm}) 
is that the coefficient functions of descendants are related to those of twist-two operators by conformal symmetry, schematically
\begin{align}
   A_N^{\mu_1\ldots \mu_N} & \stackrel{O(4,2)}{\mapsto} \{B_N,C_N, D_N,\ldots\}^{\mu_1\ldots \mu_N}, 
\end{align}
and do not need to be calculated explicitly. This approach was used to calculate 
the twist-four kinematic corrections \cite{Braun:2012hq} used in ~\cite{Braun:2014sta}. 
The extension of these results up to twist-six accuracy for scalar targets is reported in \cite{Braun:2022qly}. 
Possible applications beyond DVCS include, e.g., the studies of $t$-channel processes like $\gamma^\ast\gamma \to \pi\pi$, see 
\cite{Lorce:2022tiq}.

\section{Beyond the twist expansion}

Not all power corrections $1/Q^k$ can be obtained from the expansion near the light cone. An instructive example \cite{Musatov:1997pu} is provided by
a calculation of the pion transition form factor $\gamma^\ast \to \gamma\pi$ is the Brodsky-Lepage formalism~\cite{Lepage:1980fj}. Using a simple model 
for the quark-antiquark component of the pion light-front wave function
\begin{align}
   \Psi^{\bar q q}(x,k_\perp^2) \sim \phi_\pi(x) \exp\Big[- \frac{k_\perp^2}{2 \sigma x\bar x}\Big],\qquad \bar x = 1-x\,,
\end{align}
where $\sigma = \mathcal{O}(\Lambda_{\text QCD}^2)$ is a nonperturbative width parameter, one obtains \cite{Musatov:1997pu}
\begin{align}
 F^{\bar q q}_{\gamma^\ast\gamma\pi}(Q^2) 
= \frac{\sqrt{2}f_\pi}{3 Q^2 } \int_0^1 \frac{dx}{x} \phi_\pi(x) \Big[1 - {\exp\Big(-\frac{xQ^2}{2\bar x \sigma}} \Big) \Big],
\label{MusRad}
\end{align}
so that  for $Q^2\to\infty$ 
\begin{align}
 F_{\gamma^\ast\gamma\pi}(Q^2) 
= \frac{\sqrt{2}f_\pi}{3 Q^2 } \int_0^1 \frac{dx}{x} \phi_\pi(x) +\underbrace{\frac{1}{Q^4}[\text{higher Fock states}]}_{\text{higher twists}}
{\underbrace{-\frac{4\sqrt{2}f_\pi \sigma}{Q^4}}_{\text{end-point (Feynman)}}}.    
\end{align}
The first term in this expression is the usual pQCD result at leading power~\cite{Lepage:1980fj}, the second (higher-twist) term comes from 
the contributions of higher Fock states, and the last term arises from the exponential correction in Eq.~(\ref{MusRad}) 
under the assumption that $\phi_\pi(x)_{x\to 1} = \mathcal{O}(1-x)$.  This last contribution originates from the end-point region 
$1-x = \mathcal{O}(\sigma/Q^2)$ and large transverse distances between the quark and the antiquark in the pion, $|x_\perp^2| \sim 1/\sigma \sim 1/\Lambda_{\text QCD}^2$,
 so that it is not seen to any order in the light-cone (twist) expansion. Such terms arise naturally in models and are usually referred to as end-point contributions.
They can be present already at leading power and invalidate collinear factorization theorems, e.g., for the $B\to\pi\ell\bar\nu_\ell$ form factor. 

\begin{figure}[t]
\begin{center}
\begin{minipage}[c]{4cm}{\includegraphics[width=0.85\textwidth, clip = true]{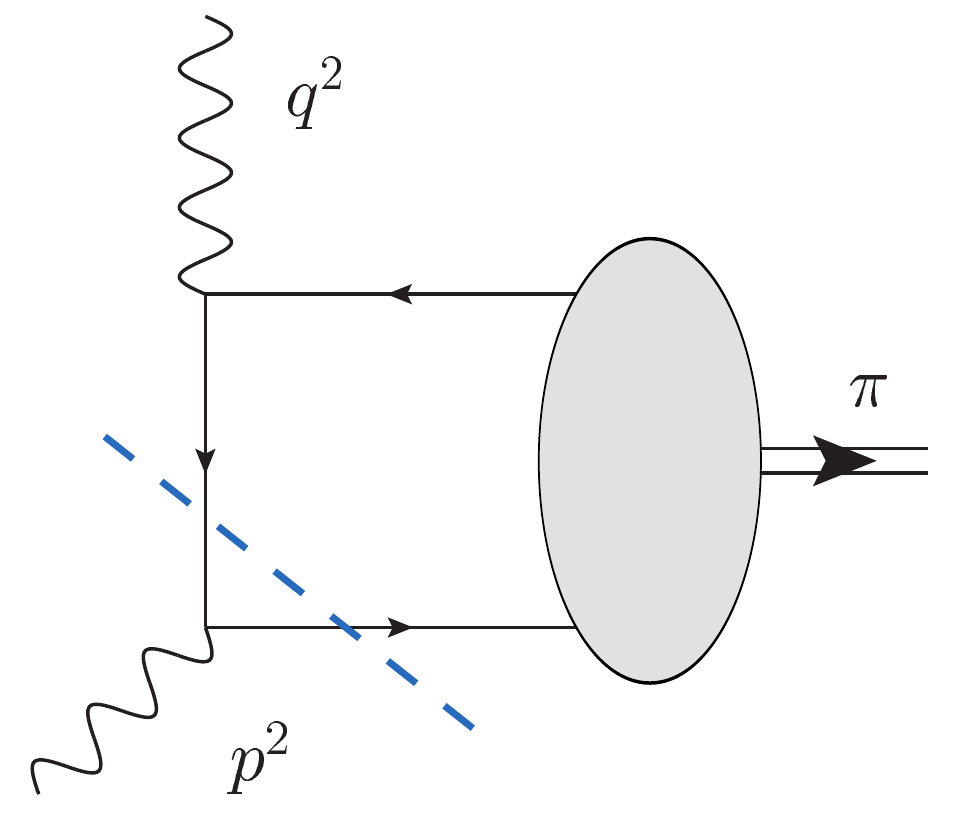}}\end{minipage}
\hspace*{0.2cm}
\begin{minipage}[c]{8cm}{\vspace*{0.6cm}\includegraphics[width=0.85\textwidth, clip = true]{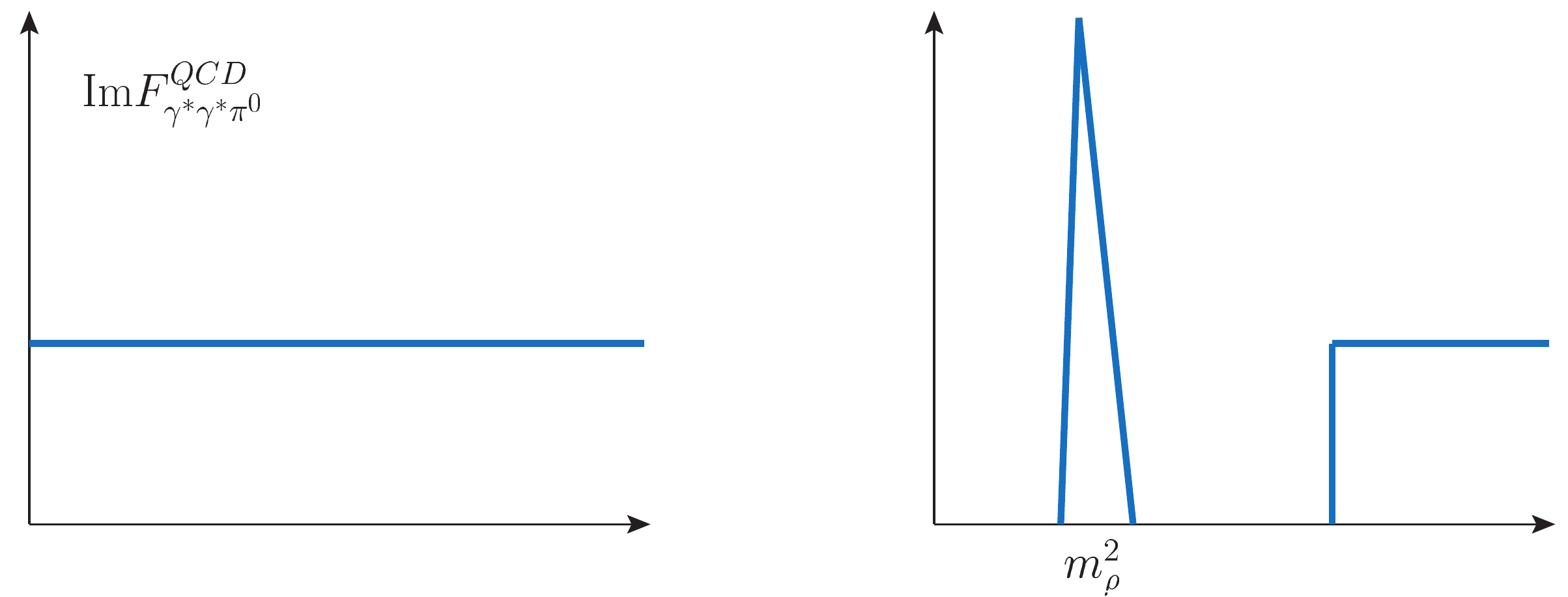}}\end{minipage}
\end{center}
\caption{A schematic representation of the LCSR technique: Consider a dispersion relation 
for the form factor with two nonzero photon virtualities (left) and modify the perturbative spectral density (middle) by 
the resonance + continuum model with the parameters fixed by duality.}
\end{figure}

End-point contributions present a new type of nonperturbative corrections that should be taken into account. 
They have to be properly defined and separated from the ``usual'' higher-twist corrections to prevent double counting. 
A common theory approach to estimate such contributions are the light-cone sum rules 
(LCSRs)~\cite{Balitsky:1986st,Balitsky:1989ry,Braun:1988qv,Chernyak:1990ag} based on dispersion relations and quark-hadron duality. 
The simplest example is provided, again, by the pion transition form factor $\gamma^\ast \to \gamma\pi$~\cite{Khodjamirian:1997tk}.
The idea is to consider a more general process with two virtual photons, $p^2<0$, $q^2 <0$,   
write a dispersion relation in $p^2$  and replace the calculated in QCD perturbative spectral density  by 
the resonance + continuum model with the parameters fixed by duality; put $p^2\to 0$ at the end. 
For the state-of-the-art calculations of the $\gamma^\ast \to \gamma\pi$ form factor in this technique 
see~\cite{Agaev:2010aq,Mikhailov:2016klg,Wang:2017ijn,Shen:2019zvh}.

The LCSR approach proves to be very flexible. It has several modifications and has been applied to many exclusive reactions, 
most notably to weak B-decays, see e.g.~\cite{Ball:1998kk,Ball:2004ye,Ball:2004rg,Khodjamirian:2010vf,Khodjamirian:2012rm}.
This is a large field of research that cannot be covered in this talk.

\section{Brief summary and further developments}

I have given a brief overview of the research on higher twist corrections in QCD that arise from subleading contributions in
light-cone dominated hard processes. The subject is very broad and the choice of topics in this review is of course influenced 
by my own work and experience. I apologize for being unable to mention many important contributions; the bibliography 
would explode if I tried. The take-home message from this report is that basic theory of higher-twist corrections is in a good shape,
but phenomenology is lagging behind. Am obvious reason is that higher twist effects are small in most situations and are difficult 
to disentangle unambiguously from leading twist contributions. Twist-three contributions are special in this respect since 
they can be directly related to observables, at least in principle. With extremely high statistical precision of experimental data 
expected from the new generation of particle accelerators, most notably the EIC, determination of the twist-three correlation 
functions in the nucleon can be achievable, albeit a challenging task.  

In conclusion, I want to mention briefly two more research directions that are currently receiving increasing attention.
The first one concerns power corrections in the formalism of TMD factorization. The motivation for this
research comes from the desire to be able to describe $p_T$-dependent observables in semi-inclusive reactions 
(where $p_T$ could be, e.g., the hadron transverse momentum in SIDIS or the transverse momentum of a Drell-Yan pair)
in the whole kinematic region from small $p_t$ where TMD factorization is applicable, to large $p_T$ that can be 
described in collinear factorization. At EIC, the regions of applicability of these approaches do not overlap, 
so one needs to extend them including power suppressed effects. There are several recent attempts to do this from the TMD side,
see \cite{Balitsky:2017gis,Vladimirov:2021hdn,Ebert:2021jhy}. This generalization is not straightforward, e.g., the structure of 
rapidity divergences in twist-three TMDs proves to be nontrivial \cite{Rodini:2022wki}.    

The second topic concerns going beyond leading power in jet observables and related quantities. This environment is specific 
in that all ingredients in factorization formulas are perturbatively calculable (if one stays at the parton level), so that
a proliferation of  distributions at subleading  powers is not of concern. Soft-collinear effective theory (SCET) offers a natural
framework for these studies, and in the last years this  field was very active: computing power corrections at fixed order,
subleading power regularization and resummation, resummation of large logarithms and the role of Glauber regions was addressed.
There exists a large literature on this subject, 
see e.g. \cite{Moult:2016fqy,Moult:2017xpp,Chang:2017atu,Beneke:2017ztn,Ebert:2018lzn,Liu:2020tzd,Liu:2020wbn}.
Threshold resummation beyond leading power is another relevant theme in LHC and EIC context, see \cite{Beneke:2022zkz} and references therein. 
\section*{Acknowledgments}

This work was supported in part by the Research Unit FOR2926 funded by the Deutsche Forschungsgemeinschaft
(DFG, German Research Foundation) under grant 409651613.

\bibliography{references}

\begin{thebibliography}{90}

\bibitem{Gross:1971wn}
D.J. Gross, S.B. Treiman, Phys. Rev. D \textbf{4}, 1059 (1971)

\bibitem{Brandt:1970kg}
R.A. Brandt, G.~Preparata, Nucl. Phys. B \textbf{27}, 541 (1971)

\bibitem{Kogut:1969xa}
J.B. Kogut, D.E. Soper, Phys. Rev. D \textbf{1}, 2901 (1970)

\bibitem{Jaffe:1991ra}
R.L. Jaffe, X.D. Ji, Nucl. Phys. B \textbf{375}, 527 (1992)

\bibitem{Wandzura:1977qf}
S.~Wandzura, F.~Wilczek, Phys. Lett. B \textbf{72}, 195 (1977)

\bibitem{AbdulKhalek:2021gbh}
R.~Abdul~Khalek et~al., Nucl. Phys. A \textbf{1026}, 122447 (2022),
  \texttt{2103.05419}

\bibitem{Politzer:1980me}
H.D. Politzer, Nucl. Phys. B \textbf{172}, 349 (1980)

\bibitem{Braun:1989iv}
V.M. Braun, I.E. Filyanov, Z. Phys. C \textbf{48}, 239 (1990)

\bibitem{Shuryak:1981kj}
E.V. Shuryak, A.I. Vainshtein, Nucl. Phys. B \textbf{199}, 451 (1982)

\bibitem{Shuryak:1981pi}
E.V. Shuryak, A.I. Vainshtein, Nucl. Phys. B \textbf{201}, 141 (1982)

\bibitem{Jaffe:1982pm}
R.L. Jaffe, M.~Soldate, Phys. Rev. D \textbf{26}, 49 (1982)

\bibitem{Jaffe:1983hp}
R.L. Jaffe, Nucl. Phys. B \textbf{229}, 205 (1983)

\bibitem{Bukhvostov:1984rns}
A.P. Bukhvostov, E.A. Kuraev, L.N. Lipatov, Sov. Phys. JETP \textbf{60}, 22
  (1984)

\bibitem{Bukhvostov:1985rn}
A.P. Bukhvostov, G.V. Frolov, L.N. Lipatov, E.A. Kuraev, Nucl. Phys. B
  \textbf{258}, 601 (1985)

\bibitem{Ellis:1982cd}
R.K. Ellis, W.~Furmanski, R.~Petronzio, Nucl. Phys. B \textbf{212}, 29 (1983)

\bibitem{Anikin:2009bf}
I.V. Anikin, D.Y. Ivanov, B.~Pire, L.~Szymanowski, S.~Wallon, Nucl. Phys. B
  \textbf{828}, 1 (2010), \texttt{0909.4090}

\bibitem{Braun:2008ia}
V.M. Braun, A.N. Manashov, J.~Rohrwild, Nucl. Phys. B \textbf{807}, 89 (2009),
  \texttt{0806.2531}

\bibitem{Braun:2009vc}
V.M. Braun, A.N. Manashov, J.~Rohrwild, Nucl. Phys. B \textbf{826}, 235 (2010),
  \texttt{0908.1684}

\bibitem{Braun:1998id}
V.M. Braun, S.E. Derkachov, A.N. Manashov, Phys. Rev. Lett. \textbf{81}, 2020
  (1998), \texttt{hep-ph/9805225}

\bibitem{Belitsky:2004cz}
A.V. Belitsky, V.M. Braun, A.S. Gorsky, G.P. Korchemsky, Int. J. Mod. Phys. A
  \textbf{19}, 4715 (2004), \texttt{hep-th/0407232}

\bibitem{Braun:2011aw}
V.M. Braun, T.~Lautenschlager, A.N. Manashov, B.~Pirnay, Phys. Rev. D
  \textbf{83}, 094023 (2011), \texttt{1103.1269}

\bibitem{Balitsky:1987bk}
I.I. Balitsky, V.M. Braun, Nucl. Phys. B \textbf{311}, 541 (1989)

\bibitem{Ji:1990br}
X.D. Ji, C.h. Chou, Phys. Rev. D \textbf{42}, 3637 (1990)

\bibitem{Koike:1994st}
Y.~Koike, K.~Tanaka, Phys. Rev. D \textbf{51}, 6125 (1995),
  \texttt{hep-ph/9412310}

\bibitem{Braun:2009mi}
V.M. Braun, A.N. Manashov, B.~Pirnay, Phys. Rev. D \textbf{80}, 114002 (2009),
  [Erratum: Phys.Rev.D 86, 119902 (2012)], \texttt{0909.3410}

\bibitem{Qiu:1991pp}
J.w. Qiu, G.F. Sterman, Phys. Rev. Lett. \textbf{67}, 2264 (1991)

\bibitem{Gockeler:2005vw}
M.~Gockeler, R.~Horsley, D.~Pleiter, P.E.L. Rakow, A.~Schafer, G.~Schierholz,
  H.~Stuben, J.M. Zanotti, Phys. Rev. D \textbf{72}, 054507 (2005),
  \texttt{hep-lat/0506017}

\bibitem{Bhattacharya:2020cen}
S.~Bhattacharya, K.~Cichy, M.~Constantinou, A.~Metz, A.~Scapellato,
  F.~Steffens, Phys. Rev. D \textbf{102}, 111501 (2020), \texttt{2004.04130}

\bibitem{Bhattacharya:2021moj}
S.~Bhattacharya, K.~Cichy, M.~Constantinou, A.~Metz, A.~Scapellato,
  F.~Steffens, Phys. Rev. D \textbf{104}, 114510 (2021), \texttt{2107.02574}

\bibitem{Burger:2021knd}
S.~B\"urger, T.~Wurm, M.~L\"offler, M.~G\"ockeler, G.~Bali, S.~Collins,
  A.~Sch\"afer, A.~Sternbeck (RQCD), Phys. Rev. D \textbf{105}, 054504 (2022),
  \texttt{2111.08306}

\bibitem{Braun:2021aon}
V.M. Braun, Y.~Ji, A.~Vladimirov, JHEP \textbf{05}, 086 (2021),
  \texttt{2103.12105}

\bibitem{Sato:2016tuz}
N.~Sato, W.~Melnitchouk, S.E. Kuhn, J.J. Ethier, A.~Accardi (Jefferson Lab
  Angular Momentum), Phys. Rev. D \textbf{93}, 074005 (2016),
  \texttt{1601.07782}

\bibitem{Balitsky:1989jb}
I.I. Balitsky, V.M. Braun, A.V. Kolesnichenko, Phys. Lett. B \textbf{242}, 245
  (1990), [Erratum: Phys.Lett.B 318, 648 (1993)], \texttt{hep-ph/9310316}

\bibitem{Stein:1994zk}
E.~Stein, P.~Gornicki, L.~Mankiewicz, A.~Schafer, W.~Greiner, Phys. Lett. B
  \textbf{343}, 369 (1995), \texttt{hep-ph/9409212}

\bibitem{Balla:1997hf}
J.~Balla, M.V. Polyakov, C.~Weiss, Nucl. Phys. B \textbf{510}, 327 (1998),
  \texttt{hep-ph/9707515}

\bibitem{Bury:2020vhj}
M.~Bury, A.~Prokudin, A.~Vladimirov, Phys. Rev. Lett. \textbf{126}, 112002
  (2021), \texttt{2012.05135}

\bibitem{Braun:2003rp}
V.M. Braun, G.P. Korchemsky, D.~M\"uller, Prog. Part. Nucl. Phys. \textbf{51},
  311 (2003), \texttt{hep-ph/0306057}

\bibitem{Ball:1998sk}
P.~Ball, V.M. Braun, Y.~Koike, K.~Tanaka, Nucl. Phys. B \textbf{529}, 323
  (1998), \texttt{hep-ph/9802299}

\bibitem{Ball:2006wn}
P.~Ball, V.M. Braun, A.~Lenz, JHEP \textbf{05}, 004 (2006),
  \texttt{hep-ph/0603063}

\bibitem{Balitsky:1990ck}
I.I. Balitsky, V.M. Braun, Nucl. Phys. B \textbf{361}, 93 (1991)

\bibitem{Ball:1998ff}
P.~Ball, V.M. Braun, Nucl. Phys. B \textbf{543}, 201 (1999),
  \texttt{hep-ph/9810475}

\bibitem{Ball:2007zt}
P.~Ball, V.M. Braun, A.~Lenz, JHEP \textbf{08}, 090 (2007), \texttt{0707.1201}

\bibitem{Beneke:1998ui}
M.~Beneke, Phys. Rept. \textbf{317}, 1 (1999), \texttt{hep-ph/9807443}

\bibitem{Beneke:2000kc}
M.~Beneke, V.M. Braun, pp. 1719--1773 (2000), \texttt{hep-ph/0010208}

\bibitem{Dasgupta:1996hh}
M.~Dasgupta, B.R. Webber, Phys. Lett. B \textbf{382}, 273 (1996),
  \texttt{hep-ph/9604388}

\bibitem{Kataev:1997nc}
A.L. Kataev, A.V. Kotikov, G.~Parente, A.V. Sidorov, pp. 355--364 (1997),
  \texttt{hep-ph/9706534}

\bibitem{Braun:2004bu}
V.M. Braun, E.~Gardi, S.~Gottwald, Nucl. Phys. B \textbf{685}, 171 (2004),
  \texttt{hep-ph/0401158}

\bibitem{Braun:2018brg}
V.M. Braun, A.~Vladimirov, J.H. Zhang, Phys. Rev. D \textbf{99}, 014013 (2019),
  \texttt{1810.00048}

\bibitem{Scimemi:2016ffw}
I.~Scimemi, A.~Vladimirov, JHEP \textbf{03}, 002 (2017), \texttt{1609.06047}

\bibitem{Beneke:2021lkq}
M.~Beneke, Eur. Phys. J. ST \textbf{230}, 2565 (2021), \texttt{2108.04861}

\bibitem{Braun:2014sta}
V.M. Braun, A.N. Manashov, D.~M\"uller, B.M. Pirnay, Phys. Rev. D \textbf{89},
  074022 (2014), \texttt{1401.7621}

\bibitem{Braun:2014paa}
V.M. Braun, A.N. Manashov, D.~Mueller, B.~Pirnay, PoS \textbf{DIS2014}, 225
  (2014), \texttt{1407.0815}

\bibitem{Guo:2021gru}
Y.~Guo, X.~Ji, K.~Shiells, JHEP \textbf{12}, 103 (2021), \texttt{2109.10373}

\bibitem{Kumericki:2009uq}
K.~Kumeri\v{c}ki, D.~Mueller, Nucl. Phys. B \textbf{841}, 1 (2010),
  \texttt{0904.0458}

\bibitem{Nachtmann:1973mr}
O.~Nachtmann, Nucl. Phys. B \textbf{63}, 237 (1973)

\bibitem{JeffersonLabHallA:2015dwe}
M.~Defurne et~al. (Jefferson Lab Hall A), Phys. Rev. C \textbf{92}, 055202
  (2015), \texttt{1504.05453}

\bibitem{Braun:2011zr}
V.M. Braun, A.N. Manashov, Phys. Rev. Lett. \textbf{107}, 202001 (2011),
  \texttt{1108.2394}

\bibitem{Braun:2011dg}
V.M. Braun, A.N. Manashov, JHEP \textbf{01}, 085 (2012), \texttt{1111.6765}

\bibitem{Braun:2020zjm}
V.M. Braun, Y.~Ji, A.N. Manashov, JHEP \textbf{03}, 051 (2021),
  \texttt{2011.04533}

\bibitem{Braun:2012hq}
V.M. Braun, A.N. Manashov, B.~Pirnay, Phys. Rev. Lett. \textbf{109}, 242001
  (2012), \texttt{1209.2559}

\bibitem{Braun:2022qly}
V.M. Braun, Y.~Ji, A.N. Manashov (2022), \texttt{2211.04902}

\bibitem{Lorce:2022tiq}
C.~Lorc\'e, B.~Pire, Q.T. Song (2022), \texttt{2209.11140}

\bibitem{Musatov:1997pu}
I.V. Musatov, A.V. Radyushkin, Phys. Rev. D \textbf{56}, 2713 (1997),
  \texttt{hep-ph/9702443}

\bibitem{Lepage:1980fj}
G.P. Lepage, S.J. Brodsky, Phys. Rev. D \textbf{22}, 2157 (1980)

\bibitem{Balitsky:1986st}
I.I. Balitsky, V.M. Braun, A.V. Kolesnichenko, Sov. J. Nucl. Phys. \textbf{44},
  1028 (1986)

\bibitem{Balitsky:1989ry}
I.I. Balitsky, V.M. Braun, A.V. Kolesnichenko, Nucl. Phys. B \textbf{312}, 509
  (1989)

\bibitem{Braun:1988qv}
V.M. Braun, I.E. Filyanov, Z. Phys. C \textbf{44}, 157 (1989)

\bibitem{Chernyak:1990ag}
V.L. Chernyak, I.R. Zhitnitsky, Nucl. Phys. B \textbf{345}, 137 (1990)

\bibitem{Khodjamirian:1997tk}
A.~Khodjamirian, Eur. Phys. J. C \textbf{6}, 477 (1999),
  \texttt{hep-ph/9712451}

\bibitem{Agaev:2010aq}
S.S. Agaev, V.M. Braun, N.~Offen, F.A. Porkert, Phys. Rev. D \textbf{83},
  054020 (2011), \texttt{1012.4671}

\bibitem{Mikhailov:2016klg}
S.V. Mikhailov, A.V. Pimikov, N.G. Stefanis, Phys. Rev. D \textbf{93}, 114018
  (2016), \texttt{1604.06391}

\bibitem{Wang:2017ijn}
Y.M. Wang, Y.L. Shen, JHEP \textbf{12}, 037 (2017), \texttt{1706.05680}

\bibitem{Shen:2019zvh}
Y.L. Shen, J.~Gao, C.D. L\"u, Y.~Miao, Phys. Rev. D \textbf{99}, 096013 (2019),
  \texttt{1901.10259}

\bibitem{Ball:1998kk}
P.~Ball, V.M. Braun, Phys. Rev. D \textbf{58}, 094016 (1998),
  \texttt{hep-ph/9805422}

\bibitem{Ball:2004ye}
P.~Ball, R.~Zwicky, Phys. Rev. D \textbf{71}, 014015 (2005),
  \texttt{hep-ph/0406232}

\bibitem{Ball:2004rg}
P.~Ball, R.~Zwicky, Phys. Rev. D \textbf{71}, 014029 (2005),
  \texttt{hep-ph/0412079}

\bibitem{Khodjamirian:2010vf}
A.~Khodjamirian, T.~Mannel, A.A. Pivovarov, Y.M. Wang, JHEP \textbf{09}, 089
  (2010), \texttt{1006.4945}

\bibitem{Khodjamirian:2012rm}
A.~Khodjamirian, T.~Mannel, Y.M. Wang, JHEP \textbf{02}, 010 (2013),
  \texttt{1211.0234}

\bibitem{Balitsky:2017gis}
I.~Balitsky, A.~Tarasov, JHEP \textbf{05}, 150 (2018), \texttt{1712.09389}

\bibitem{Vladimirov:2021hdn}
A.~Vladimirov, V.~Moos, I.~Scimemi, JHEP \textbf{01}, 110 (2022),
  \texttt{2109.09771}

\bibitem{Ebert:2021jhy}
M.A. Ebert, A.~Gao, I.W. Stewart, JHEP \textbf{06}, 007 (2022),
  \texttt{2112.07680}

\bibitem{Rodini:2022wki}
S.~Rodini, A.~Vladimirov, JHEP \textbf{08}, 031 (2022), \texttt{2204.03856}

\bibitem{Moult:2016fqy}
I.~Moult, L.~Rothen, I.W. Stewart, F.J. Tackmann, H.X. Zhu, Phys. Rev. D
  \textbf{95}, 074023 (2017), \texttt{1612.00450}

\bibitem{Moult:2017xpp}
I.~Moult, M.P. Solon, I.W. Stewart, G.~Vita, JHEP \textbf{02}, 134 (2018),
  \texttt{1709.09174}

\bibitem{Chang:2017atu}
C.H. Chang, I.W. Stewart, G.~Vita, JHEP \textbf{04}, 041 (2018),
  \texttt{1712.04343}

\bibitem{Beneke:2017ztn}
M.~Beneke, M.~Garny, R.~Szafron, J.~Wang, JHEP \textbf{03}, 001 (2018),
  \texttt{1712.04416}

\bibitem{Ebert:2018lzn}
M.A. Ebert, I.~Moult, I.W. Stewart, F.J. Tackmann, G.~Vita, H.X. Zhu, JHEP
  \textbf{12}, 084 (2018), \texttt{1807.10764}

\bibitem{Liu:2020tzd}
Z.L. Liu, B.~Mecaj, M.~Neubert, X.~Wang, Phys. Rev. D \textbf{104}, 014004
  (2021), \texttt{2009.04456}

\bibitem{Liu:2020wbn}
Z.L. Liu, B.~Mecaj, M.~Neubert, X.~Wang, JHEP \textbf{01}, 077 (2021),
  \texttt{2009.06779}

\bibitem{Beneke:2022zkz}
M.~Beneke, M.~Garny, S.~Jaskiewicz, J.~Strohm, R.~Szafron, L.~Vernazza,
  J.~Wang, PoS \textbf{LL2022}, 068 (2022), \texttt{2207.14199}

\end{thebibliography}

\end{document}